\documentclass[prl,twocolumn,tightenlines,showpacs,amsmath]{revtex4}
\usepackage{epsfig} 

\begin{document} 

\title{Bound-bound pair production in relativistic collisions}

\author{ A.B.Voitkiv, B.Najjari and A.Di Piazza} 
\affiliation{ Max-Planck-Institut f\"ur Kernphysik, 
Saupfercheckweg 1, D-69117 Heidelberg, Germany } 


\begin{abstract} 

Electron-positron pair production is considered 
in the relativistic collision of a nucleus and 
an anti-nucleus, in which both leptons are created in bound
states of the corresponding nucleus-lepton system. 
Compared to free and bound-free pair production 
this process is shown to display a qualitatively 
different dependency both on the impact energy 
and charged of the colliding particles. 
Interestingly, at high impact energies the cross section 
for this process is found to be larger than that 
for the analogous atomic process of non-radiative 
electron capture although the latter 
does not involve the creation of new particles. 

\end{abstract} 


\maketitle 

 
One of the most fascinating predictions of 
quantum electrodynamics is the possibility 
of converting energy into matter. 
Starting with the paper by Sauter \cite{Sauter_1931} 
electron-positron pair production 
from vacuum due to the presence of external electromagnetic fields 
has been attracting the attention of different physical communities.  

Pair production has been studied theoretically 
in the presence of electromagnetic fields 
of various configurations 
(e. g., in the combination of Coulomb 
and high-energy photon fields \cite{BeHe1934},
in high-energy collisions of charged particles \cite{LL}, 
in constant and uniform fields \cite{Heisenberg_1936}, 
in slowly varying super-strong Coulomb fields \cite{Greiner_b_1985}, 
in colliding laser fields \cite{Brezin_1970},  
in crystals \cite{Baier_2005}), 
and also in the presence 
of gravitational fields \cite{Birrell_b_1984}. 

Pair production can occur with noticeable probabilities 
(i) if the external field is strong enough 
to provide an energy of the order of 
the electron rest energy $mc^2$ on a distance of the order of the electron Compton wave length $\lambda_C = \hbar/mc$ \cite{supercritical}, 
where $\hbar$ is the Planck's constant,    
(ii) or/and if the field varies in time so rapidly  
that its typical frequencies multiplied by 
$\hbar$ are (at least) of the order of $2mc^2$. 

Experimentally pair production has been explored only 
in the case of rapidly varying electromagnetic fields 
(for instance, in relativistic heavy-ion collisions, 
photon-laser collisions \cite{Burke_1997}, 
in the collision of an intense laser beam 
with a solid target \cite{Chen_2009}).   
 
\begin{figure}[t] 
\vspace{-0.45cm}
\begin{center}
\includegraphics[width=0.33\textwidth]{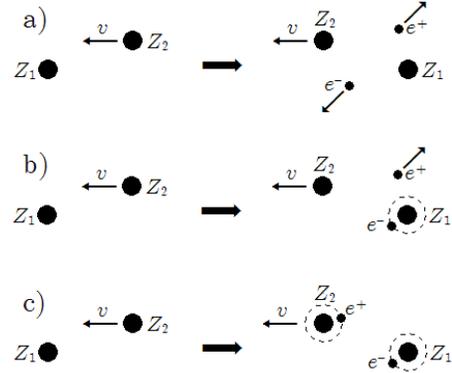}
\end{center} 
\vspace{0.25cm} 
\caption{ A sketch of different pair production 
(sub-)processes: free (a), bound-free (b) and 
bound-bound (c) pair production. }  
\label{sketch} 
\end{figure}

Landau and Lifshitz \cite{LL} 
were the first to estimate 
the cross section for pair production 
in relativistic collisions of charged particles 
in which the created electron and positron 
freely move in space after the collision 
is over (see figure \ref{sketch}a).  
Such a process is termed free  
pair production and 
it was studied in much detail in a vast amount 
of theoretical and experimental papers 
(see for recent reviews e.g. 
\cite{Baur} and also references therein). 

During the last two decades 
another kind of pair production 
process occurring 
in relativistic nuclear collisions 
has attracted much attention  
(see e.g. \cite{Anh-Gould}-\cite{Croth} 
and references therein).   
In contrast to free pair production,  
in this process the electron is created 
in a bound state with one of 
the colliding nuclei (see figure \ref{sketch}b).  

When the colliding nuclei possess 
charges of different signs, 
yet another pair production 
process becomes possible in which 
not only the electron but also the positron 
are created in a bound state (see figure \ref{sketch}c).  
Below we shall call this process bound-bound pair  production. 
Compared to the free and bound-free cases,  
bound-bound pair production is expected 
to have a number of interesting features; 
in particular, it has an intrinsic 
non-perturbative dependence on charges 
of \emph{both} colliding nuclei.   
This, as well as the fact that this process completes 
the picture of the basic (single-) pair production 
processes occurring in high-energy collisions 
of charged particles, makes its study of great interest. 
To our knowledge, bound-bound pair production 
has not yet been considered in the literature and 
it is the goal of this Letter to investigate this process.  

Let us consider the collision of two nuclei with  
charges $Z_1$ and $Z_2$ (say $Z_1 > 0$ and $Z_2 < 0$) \cite{at-units}. 
Impact parameter values characteristic for this process 
are of the order of $\lambda_C$  
and, thus, are much larger than the nuclear size.  
Therefore, one can treat the nuclei as point-like  particles. 
Since we consider high impact energies one can also 
assume that the initial velocities 
of these particles are not changed in the collision.

Our consideration of the bound-bound pair production 
will be based on the semi-classical approximation 
in which only the light particles 
(electron and positron) are considered using quantum 
theory while the heavy charges 
$Z_1$ and $Z_2$ are regarded as classical particles. 
We shall employ the rest frame of 
the charge $Z_1$ as our reference frame. 
We take the position of this charge 
as the origin and assume that in this frame 
the charge $Z_2$ moves along a straight-line 
classical trajectory 
${\bf R}(t)={\bf b} + {\bf v} t$, 
where ${\bf b}=(b_x,b_y,0)$ is 
the impact parameter, ${\bf v}=(0,0,v)$ 
is the collision velocity and $t$ is the time.   
 
In order to obtain first results 
on bound-bound pair production  
it suffices to employ 
a theoretical approach which is based 
on the simplest form of 
the transition amplitude for this process 
given by 
\begin{eqnarray}
a_{bb}({\bf b}) = -i \int_{-\infty}^{+\infty} dt %
\int d^3 {\bf r} \, \psi_f^{\dagger}({\bf r},t) \,   
\hat{W}({\bf r},t) \, \psi_i({\bf r},t). 
\label{ampl-b} 
\end{eqnarray} 
In this expression $\psi_i$ 
is the state of the negative-energy electron 
bound in the field of the charge $Z_2$, 
$\psi_f$ is the state of the electron bound 
in the field of the charge $Z_1$,   
${\bf r}=({\bf r}_{\perp};z)$ 
is the lepton coordinate 
and $\hat{W}$ is the interaction. 
The latter can be chosen either 
as the interaction with 
the field of the charge $Z_1$, 
\begin{eqnarray} 
\hat{W} = - \frac{Z_1}{r}   
\label{int-1} 
\end{eqnarray}
or with the field of the charge $Z_2$, 
\begin{eqnarray} 
\hat{W} = - \frac{ \gamma Z_2}{s} 
\left(1 - \frac{v}{c} \alpha_z \right),    
\label{int-2} 
\end{eqnarray}
where $\gamma = 1/\sqrt{1-v^2/c^2}$, $s = |{\bf s}|$, 
${\bf s}=( {\bf r}_{\perp} - {\bf b}; \gamma(z-vt) )$
and $\alpha_z$ is the Dirac matrix. 
In what follows we shall take the interaction 
in  the simpler form (\ref{int-1}) but 
one can easily show that  
both (\ref{int-1}) and (\ref{int-2}) 
yield identical results for 
the amplitude (\ref{ampl-b}).  

In the reference frame chosen 
the initial and final states read  
\begin{eqnarray}  
\psi_i &=& \sqrt{ \frac{1+\gamma}{2} } %
\left( 1 + \frac{v}{c} \frac{\gamma}{ 1+\gamma} \alpha_z \right) %
\nonumber \\ 
&& \times \chi_i({\bf s}) %
\exp( i \varepsilon_p \gamma (t - vz/c^2)) 
\nonumber \\ 
\psi_f  &=& \varphi_f({\bf r}) \exp(-i \varepsilon_e t). \label{el-pos-states} 
\end{eqnarray}
In (\ref{el-pos-states}) 
$\chi_i$ is the initial negative-energy 
state, $\varepsilon_p=mc^2-I_p$   
is the total energy of the positron where 
$I_p$ is its binding energy;   
both these quantities are  
given in the rest frame 
of the charge $Z_2$. 
Further, $\varphi_f$ 
is the bound state of the electron 
and $\varepsilon_e = mc^2 - I_e$ is 
its total energy with $I_e$ being 
the binding energy.  

The amplitude (\ref{ampl-b}) is written  
in the impact-parameter space. However, 
it is more convenient to calculate 
the cross section using the 
transition amplitude written in the momentum space,  
\begin{eqnarray}
S_{bb}({\bf q}_{\perp}) &=& \frac{1}{2 \pi} %
\int d^2 {\bf b} \, a_{bb}({\bf b}) \, %
\exp(i {\bf q}_{\perp} \cdot {\bf b}).  
\label{ampl-q-0}  
\end{eqnarray}
Using Eqs. (\ref{ampl-b}), (\ref{int-1}), 
(\ref{el-pos-states}) and (\ref{ampl-q-0}) 
we obtain 
\begin{equation}
\begin{split}
S_{bb}&({\bf q}_{\perp})=
i \frac{ Z_1 }{ 2 \pi v \gamma}    
\sqrt{ \frac{1+\gamma}{2} } 
\int d^3 {\bf r} \, \varphi_f^{\dagger}({\bf r}) \,  
\frac{1}{r} \exp(i {\bf q} \cdot {\bf r}) %
\\ 
& \times  %
\left( 1 + \frac{v}{c} \frac{\gamma}{ 1+\gamma} \alpha_z \right) %
\int d^3 {\bf s} \chi_i({\bf s}) %
\exp( - i {\bf q}' \cdot {\bf s}).  
\label{ampl-q}
\end{split}
\end{equation}
The quantities ${\bf q}$ and ${\bf q}'$ 
have the meaning of the momentum transfer 
as viewed in the rest frames of  
the charges $Z_1$ and $Z_2$, respectively, 
and are given by 
\begin{eqnarray} 
{\bf q} &=& \left( {\bf q}_{\perp}, 
\frac{mc^2 - I_e + (mc^2 - I_p)/\gamma}{v} \right) 
\nonumber \\ 
{\bf q}' &=& \left( {\bf q}_{\perp}, 
\frac{ mc^2 - I_p + (mc^2 - I_e)/\gamma }{v} \right).  
\label{momenta}  
\end{eqnarray} 
The total cross section for the bound-bound 
pair production reads  
\begin{eqnarray} 
\sigma_{bb} =      
\int d^2 {\bf q}_{\perp}  \, 
\mid S_{bb}({\bf q}_{\perp}) \mid^2.  
\label{cross-section}   
\end{eqnarray}

It follows from Eqs. (\ref{ampl-q})-(\ref{momenta}) 
that at asymptotically high collision energies 
the only dependence of the amplitude $S_{bb}$ 
on the collision energy is given by 
the factor $1/\sqrt{\gamma}$. 
Besides, taking into account the form 
of the bound states one can show that 
this amplitude is roughly proportional 
to $ Z_1^{5/2} \, |Z_2|^{5/2} $. 
Correspondingly, we obtain that the asymptotic form 
of the cross section for 
the bound-bound pair production is given by 
\begin{eqnarray} 
\sigma_{bb} \sim \frac{Z_1^5 \, |Z_2^5|}{\gamma}.  
\label{bb-asympt} 
\end{eqnarray} 
This dependence is significantly different 
from the corresponding ones in the case of free 
and bound-free pair productions which read 
$\sigma_f \sim Z_1^2 \, Z_2^2\log^3(\gamma)$
and $\sigma_{bf} \sim Z_1^5 \, Z_2^2\log(\gamma)$, 
respectively (see e.g. \cite{Baur,E-M}).  
Note that in $\sigma_{bf}$ 
$Z_1$ is the charge of the nucleus 
carrying away the created electron. 

\begin{figure}[t] 
\vspace{-0.45cm} 
\begin{center}
\includegraphics[width=0.42\textwidth]{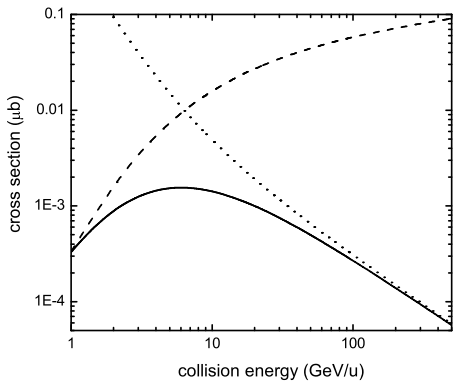}
\end{center} 
\vspace{-0.9cm} 
\caption{ Cross sections for pair production 
and electron capture given 
as a function the collision energy. 
Solid curve: p$^-$ + U$^{92+}$ $ \to $ 
$\overline{H}$(1s) + U$^{91+}$(1s). 
Dash curve: 
p$^-$ + U$^{92+}$ $ \to $ 
$\overline{H}$(1s) + e$^-$ + U$^{92+}$. 
Dot curve: H(1s) + U$^{92+}$ $ \to $ 
p$^+$ + U$^{91+}$(1s); note that 
the dot curve shows twice 
the non-radiative capture cross section. }  
\label{figure1} 
\end{figure}

In figure \ref{figure1} we show 
the cross section for the reaction 
p$^-$ + U$^{92+}$ $ \to $ $\overline{H}$(1s) + U$^{91+}$(1s) (solid curve). The dependence of 
the cross section on the impact energy is not monotonous.  
At the relatively low collision energies 
the cross section increases with the energy reaching 
a maximum at about $5$-$7$ GeV/u. With a further energy increase 
the cross section starts to decrease with an increasing slope 
and reaches its asymptotic energy dependence $\sim 1/\gamma$ 
already within the energy interval displayed in the figure. 

The cross section for the bound-bound 
pair production can be compared with 
that for the bound-free pair creation. 
We have calculated the cross section for 
the latter process 
(shown in figure \ref{figure1} by dash curve) 
treating it as 
a transition between the negative- 
and positive-energy Coulomb states 
centered on the antiptoron which 
is induced by the interaction with 
the field of the charge $Z_1$ taken 
into account in the lowest order perturbation 
theory. 
 
At relatively low impact energies both 
cross sections increase and are rather 
close in magnitude to each other.  
However, at larger impact energies 
the cross sections start 
to demonstrate qualitatively 
different behaviours and 
the difference in the magnitude 
between them increases very rapidly.  

Such an interrelation between these cross sections 
can be understood by noting the following. 
At very low collision energies the spectrum of 
the electromagnetic field generated by the colliding 
particles does not have enough high-frequency 
components necessary to create an electron-positron pair. 
As a result, the cross sections for both pair production 
processes are very small. An increase in the impact 
energy leads to an increase of the high-frequency 
component of the field and both cross sections grow rather rapidly. However, when the impact energy increases 
further the conditions for the bound-bound 
pair production begin to deteriorate. 
Indeed, the electron and positron are created 
on different nuclei and, therefore,  
the difference between their momenta 
increases with the impact energy.    
This reduces the overlap between the states 
$\psi_i$ and $\psi_f$ making bound-bound 
pair production more difficult to occur.   

This, of course, does not occur 
in bound-free pair production since both 
the leptons are created on/around the same nucleus. 
In this case when the impact energy grows 
the range of the impact parameters efficiently 
contributing to the process grows as well 
($\sim \gamma$) leading to the logarithmic 
increase in the bound-free pair 
production cross section.  
  
Bound-bound pair production can be 
viewed as a collision-induced transition between 
states of the electron with negative  
and positive total energies bound 
in the field of the charge $Z_2 < 0$ 
and charge $Z_1 > 0$, respectively. 
This is reminiscent of the  
atomic collision process 
of non-radiative electron capture 
(for a review see e.g. \cite{E-M}-\cite{Croth}) 
in which an electron initially bound 
in the atom undergoes a transition into a bound state 
in the ion: ($Z_a$ + e$^-$) + $Z_i$ $\to$ $Z_a$  + ($Z_i$ + e$^-$), 
where $Z_a$ and $Z_i$ are the charges of 
the atomic and ionic nuclei.    
Indeed, within the simplest description of 
non-radiative capture its amplitude is given by 
Eq. (\ref{ampl-b}) in which 
$\psi_i$ and $\psi_f$ are now the states 
of the electron bound in the atom and ion, 
respectively, and, therefore, 
it is of interest to compare 
the cross sections for these two processes.  

Such a comparison is presented 
in figure \ref{figure1} where dot curve    
shows twice the cross section for the reaction 
H(1s) + U$^{92+}$ $ \to $ p$^+$ + U$^{91+}$(1s)  
calculated using the simplest description 
mentioned above. At relatively low 
and intermediate collision 
energies, where the electron capture is much 
more probable than the bound-bound pair production, 
the two cross sections show 
a qualitatively different behaviour.   
However, at higher impact energies 
the cross sections approach 
each other, cross and, when the energy 
increases further, demonstrate 
exactly the same energy dependence with 
the bound-bound pair production 
cross section being a factor of $2$ 
larger. 

The factor of $2$ is of statistical origin 
reflecting the difference between the 
averaging over the projections 
of the angular momentum 
in the initial state 
in the case of electron capture  
and the corresponding summation 
in the case of pair production.   
Apart from it the cross sections 
are identical. 
This circumstance can be understood 
by taking into account the symmetry  
between these two processes and 
observing that at $ \gamma \gg 1$    
the absolute values of the momentum 
transfers in both processes become essentially the same.   
Thus, at asymptotically high impact energies 
it would be easier to capture electron and positron 
from vacuum into the corresponding bound states 
in the collision with an antiproton than to 
pick up the already existing electron 
from the atomic hydrogen. 

\begin{figure}[t] 
\vspace{-0.45cm} 
\begin{center}
\includegraphics[width=0.42\textwidth]{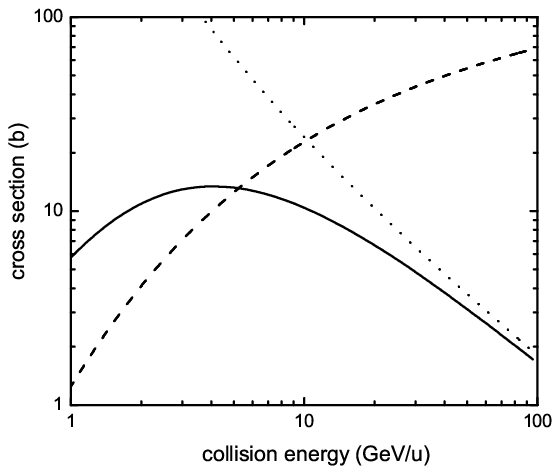}
\end{center} 
\vspace{-0.9cm} 
\caption{ The same as in figure 
\ref{figure1} but for 
$\overline{U}^{92-}$ + U$^{92+}$ $ \to $ 
$\overline{U}^{91-}$(1s) + U$^{91+}$(1s) (solid curve),  
U$^{92+}$ + U$^{92+}$ $ \to $ 
U$^{91+}$(1s) + e$^+$ + U$^{92+}$ (dash curve) 
and U$^{90+}$(1s$^2$) + U$^{92+}$ $ \to $ 
U$^{91+}$(1s) + U$^{91+}$(1s) (dot curve).}  
\label{figure2} 
\end{figure}

In order to illustrate the very strong dependence 
of the bound-bound pair production 
on the charges of the colliding particles 
in figure (\ref{figure2}) we present 
the cross section for the (hypothetical) reaction 
$\overline{U}^{92-}$ + U$^{92+}$ $ \to $ 
$\overline{U}^{91-}$(1s) + U$^{91+}$(1s). 
It is seen that compared to the case shown in 
figure (\ref{figure1}) the magnitude 
of the cross section has increased roughly by 10 orders 
with the dependence on the collision energy 
remaining basically the same.   
In figure (\ref{figure2}) the bound-bound cross 
section can also be compared with 
those for the bound-free pair production in 
U$^{92+}$ + U$^{92+}$ collisions and 
for the electron capture in collisions between 
U$^{90+}$(1s$^2$) and U$^{92+}$.     

Before we proceed to conclusions, 
two brief remarks may be appropriate. 
First, in order to calculate the cross section 
for the bound-bound pair production 
we used a simple approximation  
which is not expected to yield very precise 
cross section values. 
In case of very asymmetric collisions, 
like e.g. between antiprotons and uranium ions, 
a better estimate for this cross section can be obtained by employing an impulse-like approximation. 
Note, however, that while the cross section 
values can be somewhat changed  
if a better approximation is applied, 
we do expect that the general dependences 
on the energy and charged of the colliding particles, 
obtained using this simple treatment, 
will not be noticeably altered. 

Second, we have focused our attention 
on bound-bound pair  production in collisions 
involving a highly charged nucleus and an antiproton. 
However, this process also occurs if the antiproton 
is replaced in the collision by another particle 
with negative charge, for example by an electron 
or a muon $\mu^-$. 
In particular, since the mass of $\mu^-$ is much larger 
than that of electron/positron,  
our results obtained for collisions involving 
antiprotons are directly applicable 
to collisions with muons.  

In conclusion, we have considered 
bound-bound $e^+ e^-$ pair production in which 
both these particles are created in bound states. 
Compared to free and bound-free pair productions 
it represents a qualitatively new sub-process 
whose cross section has different dependences on the 
impact energy and charges of the colliding nuclei. 
Besides, its consideration also enables 
one to establish an interesting 
correspondence between the pair production 
and the more usual atomic collision process  
in which already existing electrons 
undergo transitions between colliding centers  
but no new particles is created.   

In non-relativistic quantum theory 
only the positive energy states exist 
and within the non-relativistic consideration 
of ion-atom (ion-ion) collisions only 
three basic atomic processes appear: 
excitation, ionization and electron capture.  
The relativistic theory adds up the negative 
energy states into consideration. 
This results in the existence 
of pair production and the corresponding 
extension of the group of the basic 
atomic collision processes to the six.   
Thus, the bound-bound pair production 
not only fills in the 'vacancy' in the set of the (single-) 
pair production processes but can also be viewed as 
completing the whole picture of 
the basic (single-lepton) atomic processes 
possible in ion-atom (ion-ion) collisions.    

The cross section for bound-bound pair production 
is very small. Therefore, the detection of  
this process in laboratory seems to be feasible 
only provided high-luminosity beams of 
heavy nuclei and antiprotons 
(or muons) are available. 
The possibility to detect bound-bound pair production 
using the future facilities at GSI (Darmstadt, Germany) 
is currently under discussion.  
 
A.D.P. thanks C.H.Keitel for helpful discussions.


\begin{thebibliography}{99}
\bibitem{supercritical} The so called critical field 
whose strength is given by $E_{cr}=m^2c^3/\hbar e=1.3\times 10^{16}\;\text{V/cm}$ 
where $e$ is the absolute value of the electron charge.
\bibitem{at-units} 
From now on atomic units are used  
unless otherwise indicated.
\end{thebibliography}
\end{document}